\documentclass[twocolumn,showpacs,preprintnumbers,amsmath,amssymb,floatfix]{revtex4}
\usepackage{graphicx}
\usepackage{psfig}
\usepackage{dcolumn}
\usepackage{bm}
\usepackage{multirow}
\usepackage{nicefrac}

\newcommand{\ket}[1]{|#1\rangle}

\newcommand{\matrEL}[3]{\langle#1|#2|#3\rangle}
\newcommand{\matrELred}[3]{\langle#1||#2||#3\rangle}

\newcommand{\elem}[2]{\ensuremath{{}^{#1}}#2}

\begin{document}


\title{Medium polarization and finite size effects 
on the superfluidity
of the inner crust of neutron stars}

\author{S.Baroni$^{a,b.c}$, F. Raimondi$^{a,b}$, F.Barranco$^{d}$, R.A.Broglia$^{a,b,e}$, A. Pastore$^{a,b}$, 
and E.Vigezzi$^{b}$} 
\affiliation{
$^a$ Dipartimento di Fisica, Universit\`a degli Studi di Milano,via Celoria 16, 20133 Milano, Italy.\\
$^b$ INFN, Sezione di Milano, via Celoria 16, 20133 Milano, Italy.\\
$^c$ TRIUMF, 4004 Wesbrook Mall, Vancouver, B.C. V6T 2A3, Canada\\
$^d$ Departamento de Fisica Aplicada III, Escuela Tecnica  Superior de Ingenieros, Camino de los Descubrimientos s/n,    41029 Sevilla, Spain.\\
$^e$ The Niels Bohr Institute, University of Copenhagen, Blegdamsvej 17, 2100 Copenhagen \O, Denmark.}

\date{\today}




\begin{abstract}

The $\elem{1}{S}_0$ pairing gap $\Delta$ associated with the inner crust of a neutron star is calculated, taking into account
the coexistence of the nuclear lattice with the sea of free neutrons (finite size effects), as well as medium polarization effects associated with the exchange of density  and spin fluctuations. Both effects are found to be important and to lead to an overall
quenching of the pairing gap.  
This result, whose quantitative value is dependent on the effective interaction used to generate the single-particle
levels, is a consequence of the balance between the attractive (repulsive) induced interaction arising from the exchange of density 
(spin) modes, balance which in turn is influenced by the presence of the protons and depends on the single-particle
structure of the system.

\end{abstract}

\maketitle

\section{Introduction}
Neutron stars are possible remnants of  supernova, an explosion 
signaling the death of a massive star when it has run out of nuclear fuel, displaying a spectacular rapid increase in brightness.

Theoretical models consistent with the experimental findings testify to the fact 
that a neutron star has a thin atmosphere and three internal regions referred to as : 
the outer crust, the inner crust, the core. In the outer crust, matter consists of 
spherical atomic nuclei and electrons.  At the bottom of this region, neutrons 
begin to drip out of nuclei, thereby producing a neutron gas between nuclei.
It is generally accepted that in the density range $0.001\rho_0 < \rho < 0.5\rho_0$, 
where $\rho_0=2.8\cdot 10^{14}$ g\;cm$^{-3}$
corresponds to nuclear saturation density, neutrons stars display a superfluid inner crust, 
consisting of a Coulomb lattice of nuclei permeated by a gas 
of free neutrons.
The superfluidity of the inner crust has important consequences concerning
different aspects of the physics of neutron stars,
%
%
%
such as its heat capacity and thus its cooling rate. 
It is also responsible  for 
a number of 
macroscopic quantum phenomena, such as quantized vortices 
(in rotating neutron stars) and quantized magnetic flux tubes 
(in magnetized neutron star cores). The interaction of vortices with nuclei forming the Coulomb lattice in the inner crust is thought to be connected with the presence of glitches in the pulsation periods of neutron stars and the postglitch relaxation of the associated period (cf. e.g. ref. \cite{Pet.Rav.:1995}).

For a consistent description of these phenomena it is necessary to take into account
finite size effects arising from the coexistence of finite nuclei and free neutrons.
Moreover, 
polarization processes can renormalize the interaction between particles
in an important way, and thus strongly influence the
superfluid properties of the system.
For example, in uniform neutron matter a strong quenching of the $\elem{1}{S}_0$ 
pairing gap arising from the polarization of the medium is generally  
predicted \cite{Lombardo}, although no consensus has been reached concerning 
the actual intensity of this effect.
Hence, superfluid properties can be, in principle, correctly described only if medium polarization processes 
(e.g. self-energy, induced interaction,
vertex corrections) are properly taken into account \cite{Cao}.

The present paper represents a step along the path which leads to this ambitious goal.
We divide the inner crust into spherical 
Wigner-Seitz (WS) cells of different densities,  and take into account both finite size effects (arising from the presence of a nucleus at the center of the cell)
and, among all possible medium polarization effects, those associated with the 
interaction induced by the exchange of medium fluctuations, 
which are expected to give the largest contribution at the low densities typical of the system under 
consideration.
We shall use a theoretical scheme which can be applied both to finite nuclei \cite{Gori2} and 
to the inner crust of a neutron star, the only essential difference between these two systems
being the value of the Fermi energy.
A preliminary study along these lines, limited to a single value of the neutron density, was published in ref. \cite{Gori1}.


\section{The mean field}

\begin{figure}[!ht]
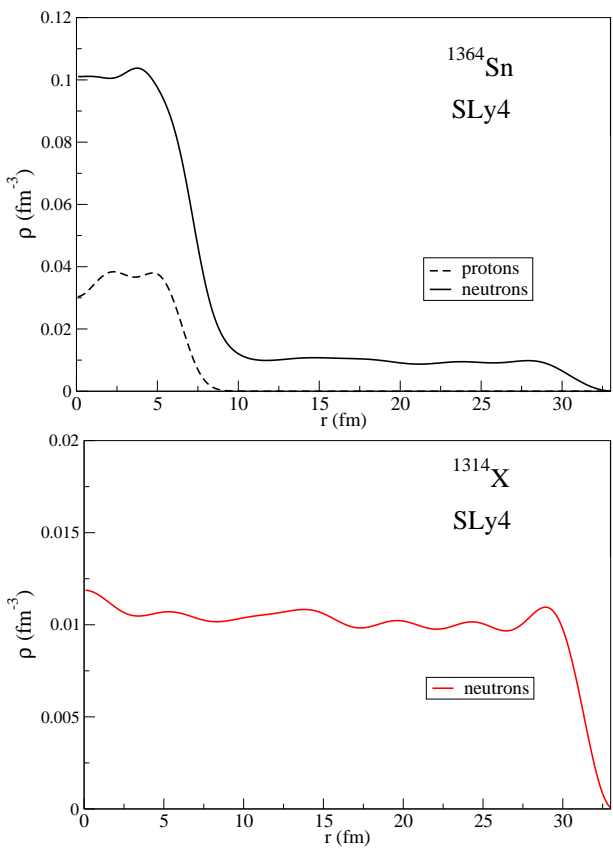

\centering
   \includegraphics*[width=0.45\textwidth,angle=0]{densita_Sn1350} \\
   \includegraphics*[width=0.45\textwidth,angle=0]{densita_X1350} 
\caption{Neutron and proton mean field densities calculated with the SLy4 interaction
in the cells \elem{1364}{Sn} 
and \elem{1314}{X} (where $X$ stands for ``no protons'') .}
\label{IND HFRPA:fig 10}
\end{figure}

We consider five regions of different densities in the inner crust
previously studied by Negele and Vautherin in the WS cell approximation \cite{NegVau}.
The number of protons, $Z$, and the radius of the cell, $R_{WS}$, have been taken from their study,
and are
listed in Table \ref{NS eye:table 10}.
We determine the single-particle states inside each cell
in the Hartree-Fock (HF) approximation, making use of either the SLy4 or the SkM$^\ast$
Skyrme interactions \cite{Cha.ea.:1998,Krivine}.
These HF calculations are analogous to those previously performed by other groups
\cite{Sandulescu,Montani}. We use
wavefunctions vanishing at the edge of the cell,
working on a grid with a mesh size of 0.1 fm.
The boundary condition affects 
the single-particle density only in 
a region of $\approx$ 3-4 fm from the edge of the cell
(see  Fig. \ref{IND HFRPA:fig 10}).  
Single-particle levels up to 100 MeV have been computed.
We have also performed calculations without protons, choosing the Fermi
energy $E_F$ so as to obtain about the same number of neutrons in the cell.
We have verified that  in the case where no protons are  considered 
we reproduce, to a good accuracy, the pairing gap found in 
infinite uniform neutron matter \cite{Pizzochero}. 
The resulting densities and the neutron numbers obtained in the various cells 
without protons are reported in Table \ref{NS eye:table 20}.  We 
also report the value of the effective mass, $m_{eff}$, associated with the effective interaction
used.
The neutron and proton density distributions calculated with the SLy4 interactions turn out to be 
quite similar to those calculated in ref. \cite{NegVau}. They   
are shown in Fig. \ref{IND HFRPA:fig 10} in the case $\rho_{n}=0.01$ fm$^{-3}$.

\begin{table}[!h]
\centering
\begin{tabular}{c|c|c|c|}
  \cline{2-4}
& & & \\
  & $Z$  & $N$      & $R_{WS}$    \\
     &    &      &    (fm) \\ 
\cline{2-4} 
  \multirow{3}{*}{SLy4}          & 40 &  458  & 42.2 \\
                          & 50 & 1048 & 35.8 \\
& 50 & 1314 & 33.2 \\
& 50 & 1740  & 27.6 \\
& 40 & 1474 & 19.6 \\
  \cline{2-4}
  \multirow{3}{*}{SkM$^*$}& 40 &  458 & 42.2 \\
                           & 50 & 1314 & 35.8 \\
                           & 50 & 1738 & 33.2 \\
  \cline{2-4}
\end{tabular}
\caption{
Proton number $Z$, neutron number $N$ and radius $R_{WS}$ for the 
WS cells containing a finite nucleus.}
\label{NS eye:table 10}

\begin{tabular}{c|c|c|c|c|c|c|c|}
  \cline{2-8}
  & $\rho_n $  & $\rho_n/\rho_0$ & $k_F$       & $E_F$   & $m_{eff}/$     &  $N$    & $R_{WS}$   \\
  & (fm$^{-3}$) &                    & (fm$^{-1}$) & (MeV)                      
& $m_{bare}$                       &      & (fm) \\ \cline{2-8}
  \multirow{5}{*}{SLy4}  &  0.0020      & 0.025             & 0.39        & 3.2   & 0.99                 &  508 & 42.2 \\
                                      &  0.007      & 0.08             & 0.58        & 7.1   & 0.97               & 1074 & 35.8 \\
                                      &  0.010       & 0.12             & 0.68        & 10.0  & 0.96                  & 1314 & 33.2 \\
                                      &  0.023       & 0.29              & 0.89        & 17.7  & 0.92                & 1760 & 27.6 \\
                                      &  0.052       & 0.65              & 1.16        & 33.5  & 0.83                & 1480 & 19.6 \\
  \cline{2-8}
  \multirow{3}{*}{SkM*}  &  0.0014      & 0.017             & 0.35        & 2.18   & 1.00               &  508 & 42.2 \\
                                      &  0.010      & 0.12             & 0.68        & 9.6   & 1.00               & 1314 & 35.8 \\
                                      &  0.023      & 0.29             & 0.89       & 16.1   & 1.00               & 1760 & 33.2 \\
  \cline{2-8}
\end{tabular}
\caption{Neutron density
$\rho_n$ (and its ratio with the neutron saturation density $\rho_0$), 
Fermi momentum $k_F$, Fermi energy $E_F$, effective mass $m_{eff}/m_{bare}$, 
neutron number $N$ and radius $R_{WS}$ for the 
WS cells without protons.
 }
\label{NS eye:table 20}
\end{table}

\section{COLLECTIVE MODES}

We have computed the excitation spectrum of the system within the framework of the 
Random Phase Approximation (RPA) in configuration space 
using the particle-hole basis states $\ket{j_pj_h^{-1},JM}$.
The residual interaction was determined
as the second derivative of the mean field energy functional,
that is, it was derived from the same two-body Skyrme force which determines
the mean field, except for the two-body Coulomb and the two-body spin-orbit interactions
which have not been included in the calculations.
RPA phonons are classified according to their total angular momentum $J$ and parity
$\pi$, while an index $\kappa=1,2,\dots$ runs over the states having the same $J^{\pi}$ values  with increasing
energy. Because  of the spin-orbit interaction, each phonon $\ket{\kappa,J^\pi}$ is
actually an admixture 
of $S=0$ and $S=1$ states. 
However, since the spin-orbit effect is very small 
for free neutrons, which dominate the RPA response,
phonons with non-natural parity are almost pure $S=1$ states, 
while natural parity states are either almost pure  $S=0$ or almost pure 
$S=1$ states.

To obtain convergence in the matrix elements of the  induced interaction, 
one needs to consider angular momenta up to about $J=30$.
For a given value of $J$, the matrix elements converge rather rapidly with the phonon 
energy, which  enters directly into the denominators of such matrix elements,
(cf. Eqs. (\ref{IND eye: eq 90}) below).
The maximum size of the particle-hole space is 4000 particle-hole configurations.
We have checked in  few cases that this is enough to achieve convergence in the
calculation of the pairing gap. 

\begin{figure}[!ht]
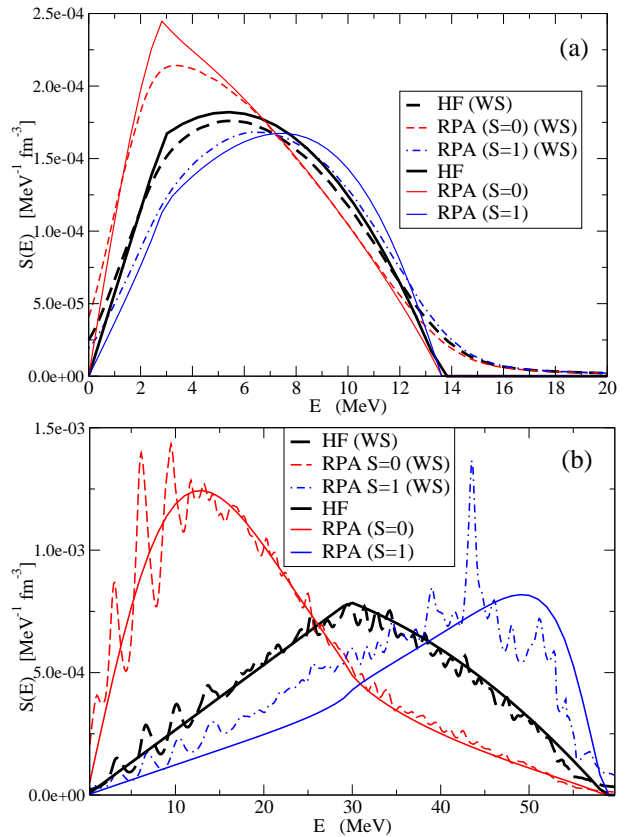

\centering
\includegraphics*[width=0.45\textwidth]{grafico_500X_100MeV_compared_sampl}\\
\includegraphics*[width=0.45\textwidth]{grafico_1500X_150MeV_compared_better_sampl}
\caption{The response functions $S$ per unit volume as a function of the energy, computed in the WS cells
at $k_{F}=$ 0.39 fm$^{-1}$ ($N= 508$) (a) and  $k_{F}=$ 1.16 fm$^{-1}$ ($N=$ 1480) (b),
are compared to the analytical results
in uniform neutron matter.
Calculations are made with  SLy4 interaction.
Both unperturbed (HF) and RPA responses are compared.
Solid lines correspond to the uniform
neutron system.
The linear momentum transferred from the external field is $q$  = 0.51 fm$^{-1}$  (a) and $q$ = 0.76 fm$^{-1}$ (b).
}
\label{NS eye: fig 20}
\end{figure}

The HF+RPA calculation without protons could be  tested against a calculation performed in neutron matter.
We have checked that in our configuration space we are able 
to reproduce  the response of the infinite uniform
neutron matter
to the external fields 
$V_{ext}(\vec{r})=e^{i\vec{q}\cdot\vec{r}}$ ($S=0$ channel) 
and $V_{ext}(\vec{r})=e^{i\vec{q}\cdot\vec{r}} \sigma_z$
($S=1$ channel).
Analytic formulae for such responses have been given in ref. \cite{Vangiai}.
The response function in the WS cells  has been evaluated according to 
\begin{equation}
\label{NS eye: eq 20}
  S(q,E)=\sum_{n=0}^{+\infty}\vert \matrEL{n}{ e^{i\vec{q}\cdot\vec{r}} \Theta_s}{0}\vert^2 L(E,E_{n}),	
\end{equation}
where $\ket{n}$ are the (mean field or RPA) excited states of the system, 
$L$ is a Lorentzian function used to smooth out the  computed discrete response,
and $\Theta_s = 1$ or $\sigma_z$  for $S=0$ and $S=1$ respectively.
Because of the spherical symmetry adopted, we have expanded  the external operator 
in multipoles and obtained the total response function adding the contributions from 
each multipolarity.
For this comparison the SLy4 parametrization has been used.
The analytical response function for uniform neutron matter is compared with the WS
cell approximation in Figs. (\ref{NS eye: fig 20}a) and (\ref{NS eye: fig 20}b)
for a neutron gas with a Fermi momentum $k_F$=0.39 fm$^{-1}$ and $k_F$=1.16 fm$^{-1}$ respectively. 
These two systems correspond to the lowest and highest density regions of the inner crust of a neutron
star (WS cells with 508 and 1480 neutrons respectively). The comparison has been
performed for a transferred momentum $q$ equal to 0.51 fm$^{-1}$ for the lower density region and
0.76 fm$^{-1}$ for the higher density region.


The overall agreement found between the numerical and analytic results 
testifies to the fact that the WS cell approximation  is a reliable tool to calculate the response
 of the free neutron gas which permeates the nuclear lattice.
We assume that the same is also true for the case of a WS cell with a nucleus in the center.
Within this context, a more detailed study of the RPA response will be given in a different paper.

\section{Induced interaction matrix elements}\label{sec: IND INT}

The matrix elements of 
the induced pairing interaction in the $\elem{1}{S}_0$ channel
\begin{equation}\label{IND eye: eq 80}
\langle v_{ind} \rangle =
   \begin{array}{c}
     \includegraphics[width=2.5cm,angle=0]{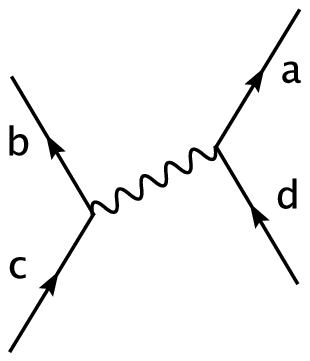}
   \end{array}
  = \matrEL{ab;00}{v_{ind}}{cd;00}, 
\end{equation}
where $\ket{ab;00}$ ($a \equiv {n_al_aj_a}$), indicates a normalized two-particle state coupled to total angular momentum zero,
are computed in Bloch-Horowitz perturbation theory \cite{Blo.Hor:1955,Dus.Lio:1971,Bes.ea:1976,Bes.ea:1977}:
\begin{eqnarray}\label{IND eye: eq 90}
  \lefteqn{\matrEL{ab;00}{v_{ind}}{cd;00}= } \nonumber \\
   & = & \sum_{int}\frac{\matrEL{ab;00}{H_{coupl}}{int}\matrEL{int}{H_{coupl}}{cd;00}}
                  {E_0 - E_{int}},\nonumber \\
		  & &
\end{eqnarray}
where $E_{int}$ is the energy of the intermediate state made by two-quasiparticles and one phonon,
\begin{equation}\label{IND eye: eq 92}
E_{int} = E_b + E_d + E_{\lambda}.  
 \end{equation}
The quasiparticle energies are denoted by $E_b$ and $E_d$, while 
$E_{\lambda}$ is the energy of the exchanged phonon and 
$E_0$ is the binding energy of a  Cooper pair.
In keeping with the results of previous studies (cf. e.g. \cite{Gori2,Bar.ea:1999,Ter.ea:2002b,Bar.ea:2004b})
we have set $E_0$ equal to $-2{\Delta_F}$, where ${\Delta_F}$ represents the value of the state-dependent pairing gap obtained solving the HFB equation, averaged over the single-particle states lying 
in the energy range
$E_F \pm 2$ MeV. 
We also computed the second time ordering (see Eq. (\ref{IND eye: eq 80})),
associated with the intermediate state of energy $E_{int} = E_a + E_c + E_{\lambda} $.


$H_{coupl}$ represents the particle-vibration coupling Hamiltonian.
It reads
\begin{equation}\label{IND eye: eq 100}
  H_{coupl}  = \sum_{mi\lambda}f^\lambda_{mi}\left(A^\dagger_{mi} O_\lambda+A_{mi} O^\dagger_\lambda   \right),
\end{equation}
with
\begin{eqnarray}\label{IND: eq 110}
  f^\lambda_{mi} = 
   \begin{array}{c}
     \includegraphics[width=1.5cm,angle=0]{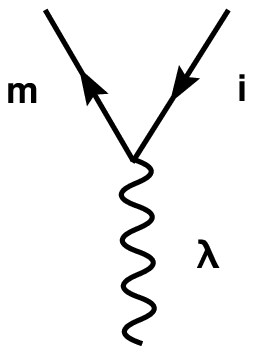}
   \end{array}
   & = & \sum_{nj}\left(\bar{v}_{mjin}X^\lambda_{nj}+\bar{v}_{mnij}Y^\lambda_{nj}  \right), \nonumber \\
   & &
\end{eqnarray}
where $A^\dagger_{mi}$ creates the particle-hole excitation $\ket{m(i)^{-1}}$, $O^\dagger_\lambda$ is the phonon
creator operator, $X^\lambda_{nj}$ and $Y^\lambda_{nj}$ are the forward and backward amplitudes of the 
phonon $\ket{\lambda}$ associated with the particle-hole component $\ket{n(j)^{-1}}$.
$\bar{v}_{mjin}$ is   the RPA antisymmetrized matrix element of the residual interaction $v_{res}$. 
We treat the latter in the Landau approximation. Thus
 \begin{eqnarray}\label{IND eye: eq 120}
  v_{res}(\vec{r},\vec{r}') & = & \left[ F_0(r)+F'_0(r)\vec{\tau}\cdot\vec{\tau}'\right]\delta(\vec{r}-\vec{r}') \nonumber \\
                            &   & + \left[G_0(r)+G'_0(r)\vec{\tau}\cdot\vec{\tau}'\right]\vec{\sigma}\cdot\vec{\sigma}'\delta(\vec{r}-\vec{r}').
                           \nonumber \\
   & &
\end{eqnarray}
Exchange terms of $v_{res}$ have not been considered.
Moreover, as far as the computation
of the induced interaction matrix
elements is concerned, the terms associated
with the momentum dependent terms of the NN effective interaction have been neglected \cite{Gori2}. 
An analysis of the influence of these terms will be presented in future work.
We shall only consider the $\tau_z \cdot \tau'_{z}$ term, in keeping with the fact that we are interested in the neutron-neutron
pairing interaction. 
In uniform neutron matter, RPA phonons are characterized by their spin: density ($S=0$) modes are associated uniquely with the spin-independent part of the interaction,
given by
$C_F(r) \equiv F_0(r) + F'_0(r)$, while 
spin ($S=1$) modes are produced by the spin-dependent part, 
$C_G(r) \equiv G_0(r) + G'_0(r)$. In the WS cell, one has to take into account also
the proton degree of freedom
($\tau_z\cdot\tau'_z$ =-1)  
contribution to the RPA response. Moreover, because of the spin-orbit interaction, each phonon is an admixture of $S=0$ and $S=1$ states. While coupling to non-natural parity modes can take place  only through the spin-dependent part of the interaction,  coupling to natural parity modes can receive contributions from both 
channels, although the spin-independent part is usually the dominant one.

\begin{figure}[!ht]
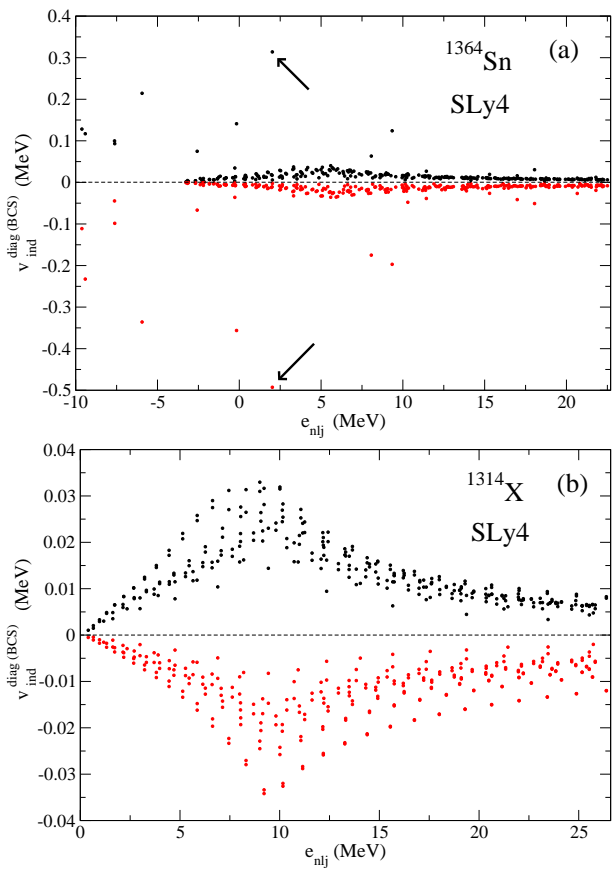

\centering
  \includegraphics*[width=0.45\textwidth,angle=0]{grafico_edm_indotta_BCS_sommati_umaretatoki_1350Sn} \\ 
  \includegraphics*[width=0.45\textwidth,angle=0]{grafico_edm_indotta_BCS_sommati_umaretatoki_1350X}
\caption{Induced pairing interaction diagonal matrix elements (in the BCS limit) for the cells 
\elem{1364}{Sn} (a) and \elem{1314}{X} (b), computed with the SLy4 effective interaction.
Positive (negative) matrix elements correspond to abnormal parity  (normal parity) modes. 
Note the different scale of the two figures. The arrows indicate the two matrix elements discussed in the text,
and associated with the transition densities shown in Fig. \ref{NS ind:fig 50}.}
\label{NS ind:fig 30}
\end{figure}

\begin{figure}[!ht]
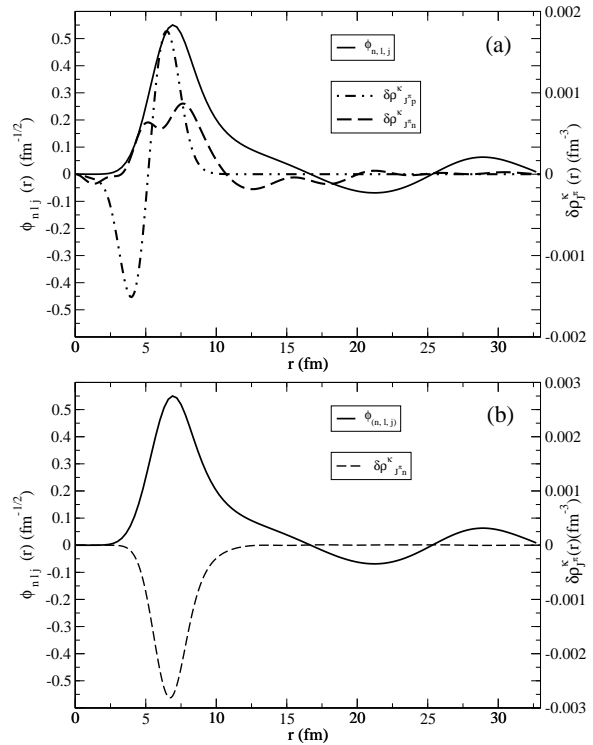

\centering
 \begin{minipage}[l]{0.90\linewidth}
   \includegraphics*[width=1.\linewidth,angle=0]{grafico_wf_3_7_15_and_TD_2_2_0_158}
  \end{minipage}
 \begin{minipage}[l]{0.90\linewidth}
   \includegraphics*[width=1.\linewidth,angle=0]{grafico_wf_3_7_15_and_TD_15_14_1_3192}
  \end{minipage}
 \caption{
(a) Proton and neutron transition densities associated with the J$^\pi=2^+$ normal parity 
 phonon with energy 4.46 MeV  for the cell \elem{1364}{Sn}.
The single-particle wavefunction $\phi_{nlj}=\phi_{3,7,15/2}$ is also shown.
(b) The same, for the  J$^\pi=15^+$ (L=14) non-normal parity phonon at  15.27 MeV.
There is no proton contribution to this transition density within our particle-hole space.}
%
\label{NS ind:fig 50}
\end{figure}

The explicit expressions for the vertices $f$ and $g$ in the $S=0$ and $S=1$ channels, 
 coupling the neutron single-particle 
states   to the $\kappa$-phonon of angular momentum and parity 
$J^{\pi}$, are

\begin{eqnarray}\label{NSind:eq40A}
   \lefteqn{
   \begin{array}{c}
     \includegraphics[width=3.0cm,angle=0]{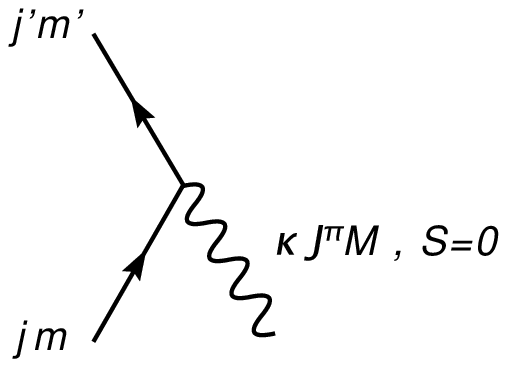}
   \end{array} = f     } \nonumber \\
    & = & \matrEL{j'm'}{H_{coupl}}{jm,\kappa J^\pi M}  \nonumber \\
    & = & 
      \matrEL{j'm'}{i^J Y_{JM}}{jm}\int dr\phi_{\nu'}(r)
[(F_0+F'_0) \delta\rho^{\kappa}_{J^\pi n}(r) + \nonumber  \\
& & (F_0-F'_0) \delta\rho^{\kappa}_{J^\pi p}(r) ]
\phi_{\nu}(r)\nonumber \\
  & &
\end{eqnarray}   
\begin{eqnarray}\label{NSind:eq40C}
  \lefteqn{
  \begin{array}{c}
     \includegraphics[width=3.0cm,angle=0]{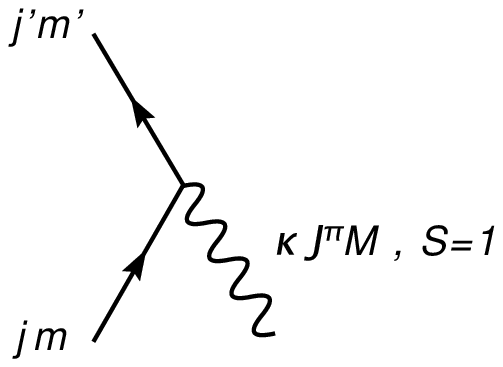}
   \end{array}
    =g  } \nonumber \\
    & = & \matrEL{j'm'}{H_{coupl}}{jm,\kappa J^\pi M}  \nonumber \\
    & = &  \sum_{L=J - 1}^{J+1} 
              \matrEL{j'm'}{i^L\left[Y_{L}(\Omega) \times \sigma\right]_{JM}}{jm} \qquad \quad \nonumber \\
    &  & 
              \int dr \phi_{\nu'}(r)
[(G_0+G'_0) \delta\rho^{\kappa}_{J^\pi L n}(r) + \nonumber \\
& & (G_0-G'_0) \delta\rho^{\kappa}_{J^\pi L p}(r) ]
\phi_{\nu}(r),  \nonumber \\
  & &
\end{eqnarray}
$\phi_{\nu}(r)/r$ is the radial single-particle wavefunction associated with
the quantum numbers $\nu \equiv \{{nlj}\}$. 
In the above expressions $\delta \rho^{\kappa}_{J^{\pi}n}$ and 
$\delta \rho^{\kappa}_{J^{\pi}p}$ are
the neutron and proton contributions
to the  transition densities associated with  the RPA modes $\ket{\kappa J^{\pi}}$. 
For the  $S=0$ channel  one finds 
\begin{eqnarray}\label{NS ind: eq 41}
     \lefteqn{ 
\delta\rho^{\kappa}_{J^\pi}(r)  } \nonumber \\
      & = &
           \Bigg[ \sum_{ph}\bigg(
                               X^\kappa_{ph}(J^\pi)+Y^\kappa_{ph}(J^\pi)
                               \bigg)
                           \phi_{\nu_h}(r)/r  \times \nonumber \\
      & &
                           \phi_{\nu_p}(r)/r \; \matrELred{j_p}{i^J Y_{J}}{j_h}
                               \frac{1}{\sqrt{2J+1}}
           \Bigg]   
,
     \nonumber \\
     & &
\end{eqnarray}
while in the $S=1$ channel we have 
\begin{eqnarray}\label{NS ind: eq 41.2}
    \lefteqn{ 
\delta\rho^{\kappa}_{J^\pi L}(r)  } \nonumber \\ 
        & = &   \Bigg[ \sum_{ph}\bigg(
                               X^\kappa_{ph}(J^\pi)-Y^\kappa_{ph}(J^\pi)
                               \bigg)\phi_{\nu_h}(r)/r \times \nonumber \\
      &  &
                               \phi_{\nu_p} (r)/r  \; \matrELred{j_p}{i^L \left[ Y_{L}(\Omega)\times \sigma \right]_J }{j_h}
                               \frac{1}{\sqrt{2J+1}}
           \Bigg].   \nonumber \\ 
         & &	   
     \nonumber \\
     & &
\end{eqnarray}
The neutron and proton contributions 
are obtained summing over the neutron or proton particle-hole states.
Since  
the sign of the denominator in the expression of the induced pairing matrix element
is always negative (see eq. \ref{IND eye: eq 90}), and since
for scattering vertices there is an additional negative sign in the $S=1$ channel 
as compared  to the $S=0$ channel \cite{Gori2},
it follows that
the exchange of vibrations 
gives rise to an attractive pairing interaction in the $S=0$ channel, and
to a repulsive interaction in the $S=1$ channel.
Hence, the  induced interaction arising from the exchange of normal parity modes involves 
in general a mixture of attractive and repulsive contributions, while for non-normal parity modes
only repulsive terms are present. 
One can show, however, that the repulsive contribution    
to the diagonal matrix elements  $\matrEL{nlj,nlj;00}{v_{ind}}{nlj,nlj;00}$ associated with normal parity modes vanishes.
These diagonal matrix elements
are shown in Fig. \ref{NS ind:fig 30}(a)  
for the case where protons are considered (\elem{1364}{Sn}), and in Fig. \ref{NS ind:fig 30}(b)
for the case without protons, (\elem{1314}{X}), corresponding to the density $\rho_{n}=0.01$ fm$^{-3}$.
They have been calculated with the SLy4 force. 
The matrix elements of the two systems are of the same order of magnitude
($\approx$0.01 MeV), except for a few, much  larger matrix elements which appear when protons are present. 
The large matrix elements  connect two-particle states based on resonant single-particle states 
which have positive energy $e_{nlj}$
and are spatially localized within the nucleus
(particularly on the nuclear 
surface).
The largest contributions  to these matrix elements arise  from collective phonons with a 
well defined surface character, both for density and spin modes.
As an example, let us focus on the diagonal matrix elements associated with
the single-particle state ($n=3,l=7,j=15/2$; $e_{3,7,15/2}= 2.02$ MeV),
whose values are equal to -0.49 MeV for normal parity, and to +0.31 MeV for non-normal parity  
(see arrows in  Fig. \ref{NS ind:fig 30}(a)). 
The transition
densities of the phonons which give rise to the most important contribution to each of these 
large matrix elements are shown in Fig. \ref{NS ind:fig 50}, 
together with the single-particle wavefunction $\phi_{3,7,15/2}$.

\begin{figure}[!ht]
\centering
\includegraphics*[width=0.45\textwidth,angle=0]{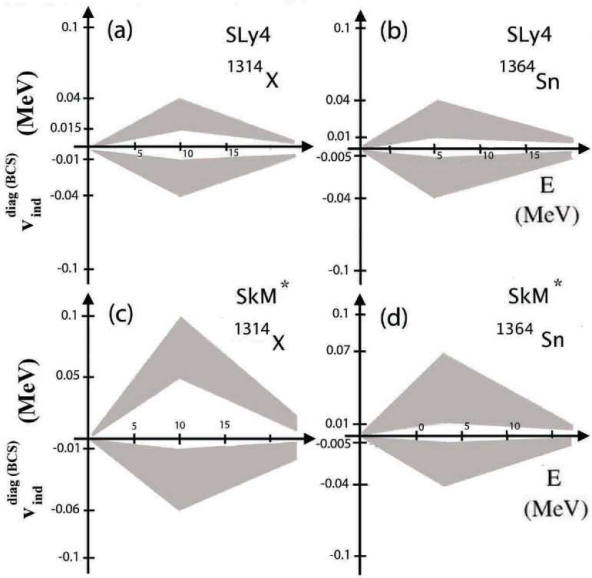}
\caption{Schematic representation of the BCS diagonal induced interaction matrix elements
for the two cells \elem{1314}{X} (left panels) and \elem{1364}{Sn} (right panels) obtained making use of SLy4
(upper panels) and SkM$^\ast$ (lower panels) effective interactions. The grey area represents the range
of the computed matrix elements (panel (a) corresponds to Fig. (\ref{NS ind:fig 30}b), while panel (b) corresponds
to Fig. (\ref{NS ind:fig 30}a) except for the few, larger and scattered matrix elements).}
\label{schematic matrix elements}
\end{figure}

\begin{figure}[!ht]
\centering
\includegraphics*[width=0.45\textwidth,angle=0]{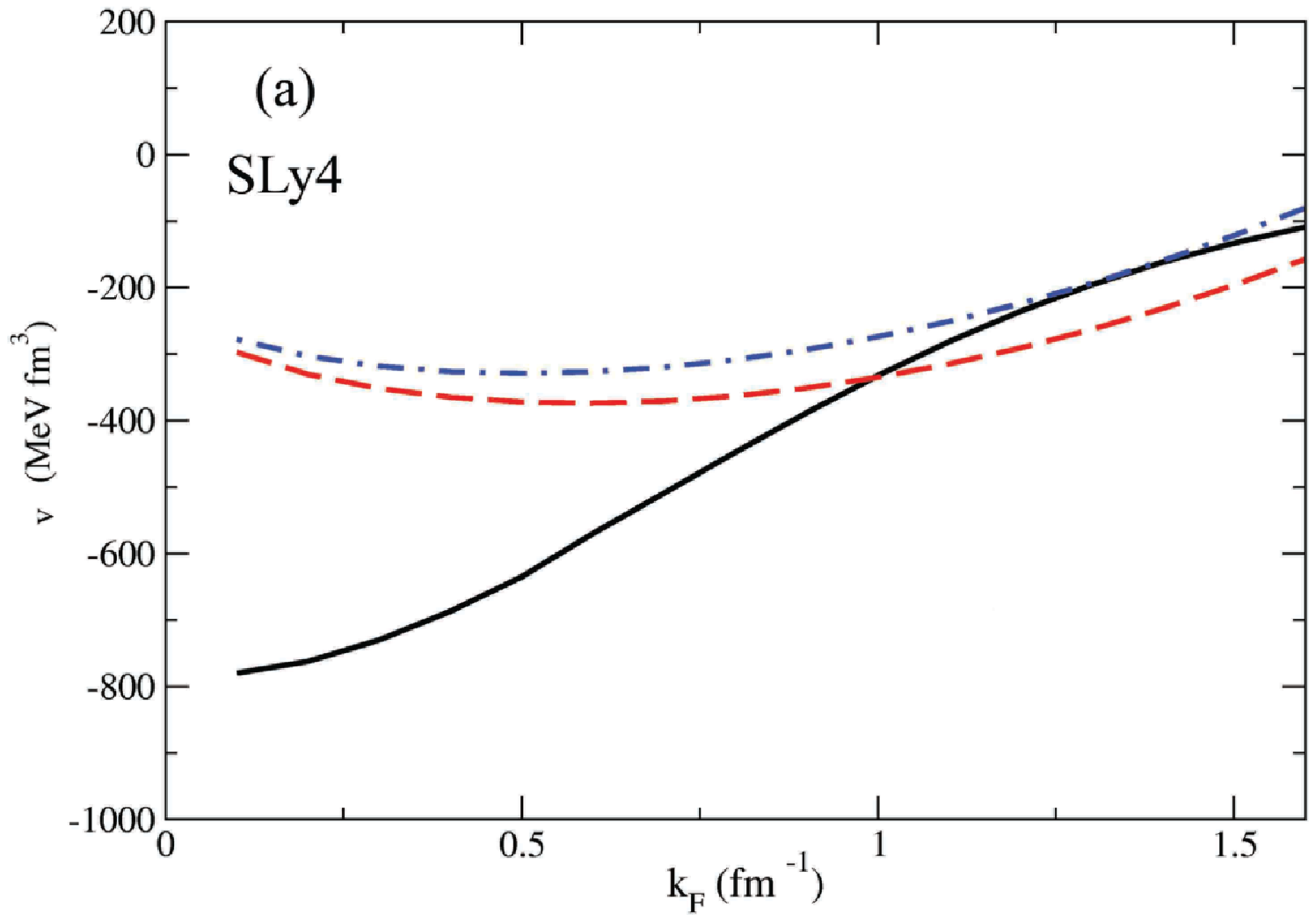}
\includegraphics*[width=0.45\textwidth,angle=0]{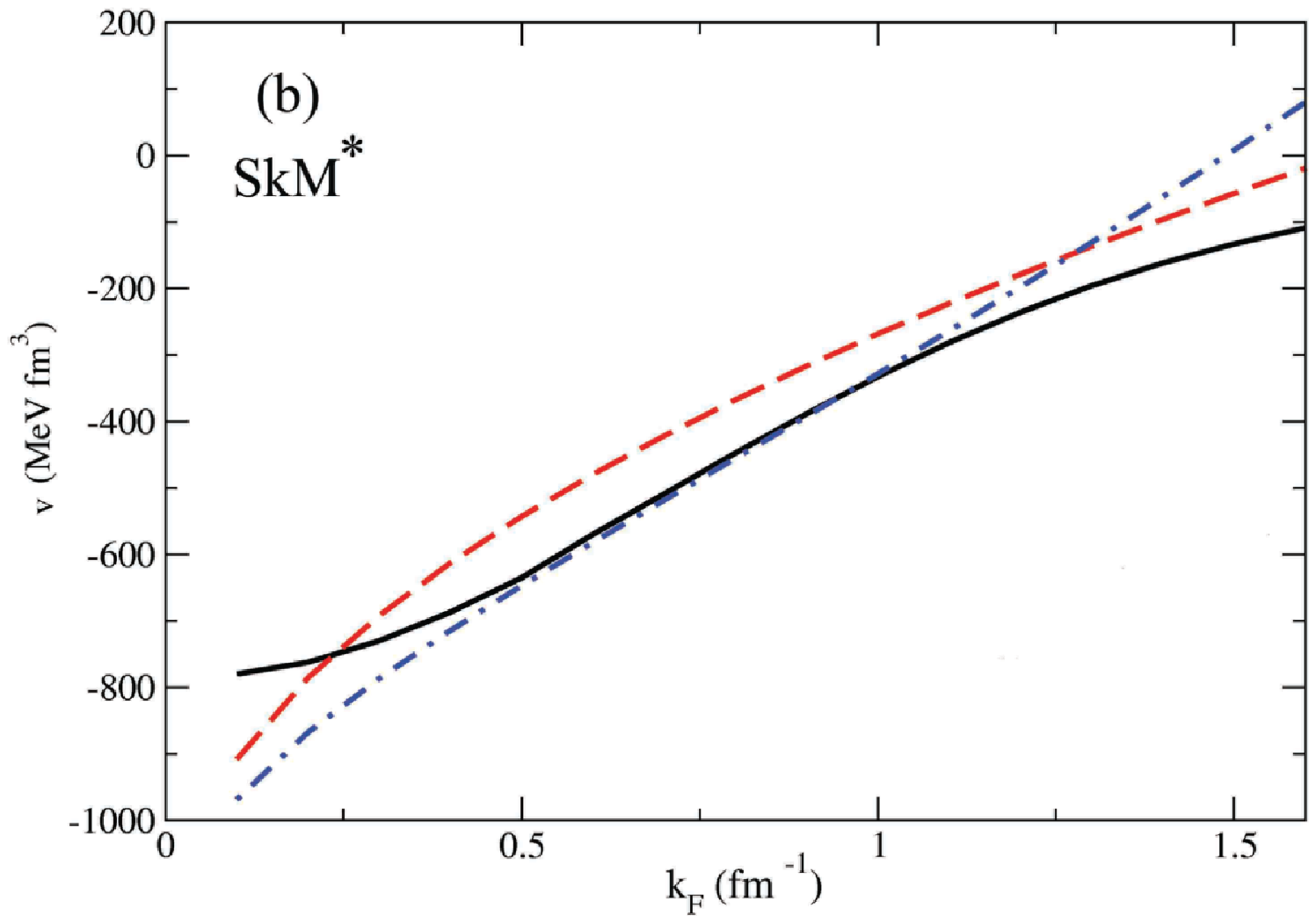}
\caption{Diagonal matrix elements of the Gogny interaction at the Fermi level (solid line) compared to the residual
interaction form factors 
$C_F$ (dashed lines) and $C_G$ (dash-dotted lines)
for the uniform neutron matter
when the SLy4 (a) and the SkM$^\ast$ (b) Skyrme interactions are employed.
(cf. Eq. (\ref{IND eye: eq 120}))}
\label{gogny vs residual}
\end{figure}

The matrix elements obtained making use of the SkM$^\ast$ interaction are qualitatively similar,
but  there are significant quantitative differences.
 A schematic plot of the diagonal induced interaction
matrix elements obtained with the two Skyrme forces used in the present paper 
is shown in Fig. \ref{schematic matrix elements}.
It can be seen that the matrix elements associated with the SkM* force are larger,
in absolute value, than those associated with the Sly4 force. Moreover,   
the spin modes have matrix elements larger (in absolute value)
than the density  modes, while they are about the same for SLy4.
This can be understood considering the Landau parameters factors $C_F$ and $C_G$ as a function
of density in uniform neutron matter. As it can be seen in Fig. \ref{gogny vs residual}, in particular at low density, the parameters associated with the 
SkM* have much larger values (in absolute value); moreover $C_G$ is larger than $C_F$, while 
the opposite is true for SLy4. 

%

\section{The pairing gap}
The pairing matrix elements of the bare and induced interaction were used in  a HFB 
calculation to study the superfluid properties of the system in the
\elem{1}{S}$_0$ channel. The HFB calculation is self-consistent only in the particle-particle channel,
while single-particle eigenstates and eigenvalues are kept fixed and equal to those computed 
in the HF calculation used to produce the mean field discussed previously.

We first consider pairing correlations at the mean field level,
employing the Gogny interaction \cite{Dec.Gog:1980} as the  bare interaction $v_{bare}$.
This is rather well  justified as long as the neutron density is smaller than, or of the order of
0.01 fm$^{-3}$ ($k_F = 0.7 $ fm$^{-1}$), because the pairing gaps obtained with the Gogny or the
bare interaction are quite similar in that density range. Instead for
higher neutron densities and in the nuclear interior the Gogny interaction produces
somewhat larger gaps. While this difference is important for a quantitative study of the pairing gap 
both in infinite and in finite systems \cite{Schuck}, it may be neglected  in the present context,
where a qualitative discussion of the role  pairing interaction plays in the neutron star crust is the
main issue.

The state dependent pairing gap $\Delta_{nn'lj}$ associated with the two-particle state
$\ket{nlj,n'lj}$ is given by 
\begin{eqnarray}\label{NS HFB: eq 10}
\Delta_{n_1n_1'l_1j_1} &=& -\frac{1}{2}\sum_{n_2n_2'l_2j_2}\sqrt{\frac{2j_2+1}{2j_1+1}}
                  \left( \sum_q U^q_{n_2l_2j_2}V^q_{n_2'l_2j_2}  \right) \nonumber \\
&  &\matrEL{n_1l_1j_1,n'l_1j_1;00}{v_{tot}}{n_2l_2j_2,n_2'l_2j_2;00}, \nonumber \\
    & &
\end{eqnarray}
where $U^q_{nlj}$ and $V^q_{nlj}$ are the quasiparticle amplitudes  obtained solving the HFB 
equations in the pairing channel, and $v_{tot} = v_{bare}+ v_{ind}$.
We have taken into account the single-particle levels contained  in an energy window going from $E_F \pm 9$  MeV
for the lowest density  WS cell to $E_F \pm 20$ MeV for the highest density WS cell.
Increasing the window by 1 MeV or including one more multipolarity 
changes the pairing gap by less than 0.01 MeV in all cases.   

\subsection{Cells without protons}

\begin{figure}[!ht]
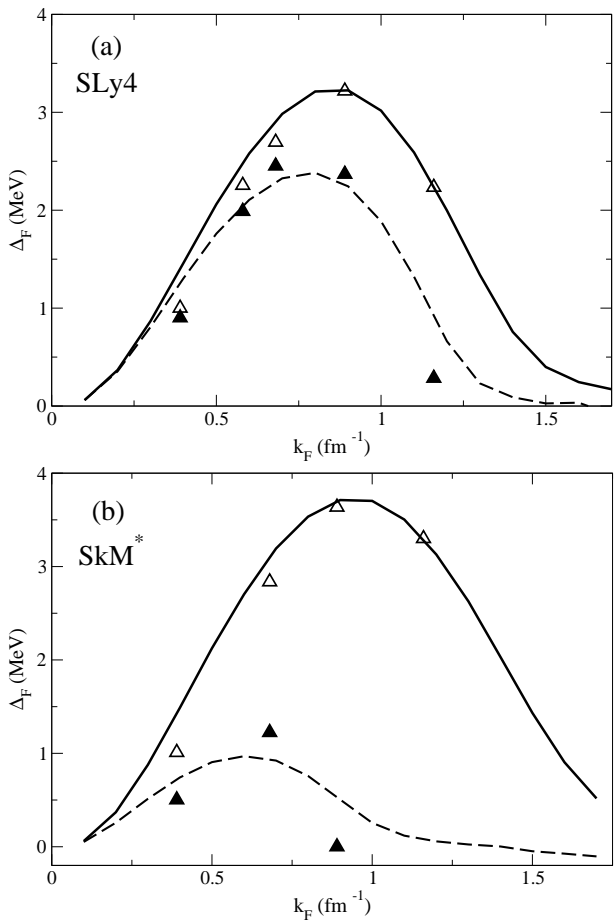

\centering
\includegraphics*[width=0.45\textwidth,angle=0]{pairing_gap_vs_Schulze_SLy4.eps}
\includegraphics*[width=0.45\textwidth,angle=0]{pairing_gap_vs_Schulze_SkmS.eps}
\caption{Pairing gap calculated at the Fermi energy for the WS cells without protons 
employing
the SLy4 (a) and the SkM$^\ast$ (b) Skyrme interactions. Empty and filled triangles refer to the 
results of the microscopic calculation with the Gogny interaction only
and with the Gogny plus the induced interactions respectively. The solid line corresponds
to a BCS calculation of the pairing gap in neutron matter, while the dashed line
corresponds
to a simpler calculation (see Appendix A) which employs a Landau approximation of the residual interaction both for RPA and for induced matrix elements.}
\label{IND eye: fig 60}
\end{figure}

The values of $\Delta_F$ obtained with and without the induced interaction on top of 
the Gogny interaction, and calculated in the  WS cells without protons,
are shown by filled  and open triangles in Fig. \ref{IND eye: fig 60}. 
It is seen that in uniform neutron matter the induced interaction   quenches the value of the 
pairing gap, and that the suppression is stronger at large densities.

When the SLy4 interaction is used, the induced interaction 
reduces the maximum value of the pairing  field by  25\%, 
bringing it from about 3.2 MeV to about  2.4 MeV. 
Moreover the gap goes rapidly to zero around  $k_F \approx$ 1.2 fm$^{-1}$,
compared to the progressive decrease occurring when only the Gogny interaction
is acting. 
The effects of the induced interaction associated with the SkM$^\ast$ case, are even more drastic.
In this case the pairing gap goes to zero already at $k_F \approx 0.9$ fm$^{-1}$ and the gap
shows a maximum value of about 1.2 MeV.

The results reported above can be checked and better understood, calculating the
pairing gap in infinite uniform neutron matter with and without the induced interaction
using the simple expressions collected in Appendix A, which lead to 
the curves shown in Fig. \ref{IND eye: fig 60}.
The agreement with the much more involved, discrete calculation performed in the WS cells is
quite satisfactory, also  considering that some approximations have been introduced 
in the expressions used in  infinite matter.
The stronger screening effect obtained with the SkM* interaction is related to the
different balance between spin and density modes already discussed in the previous section
in relation with Fig. \ref{gogny vs residual}.

\subsection{Cells with protons} 

\begin{figure}[!ht]
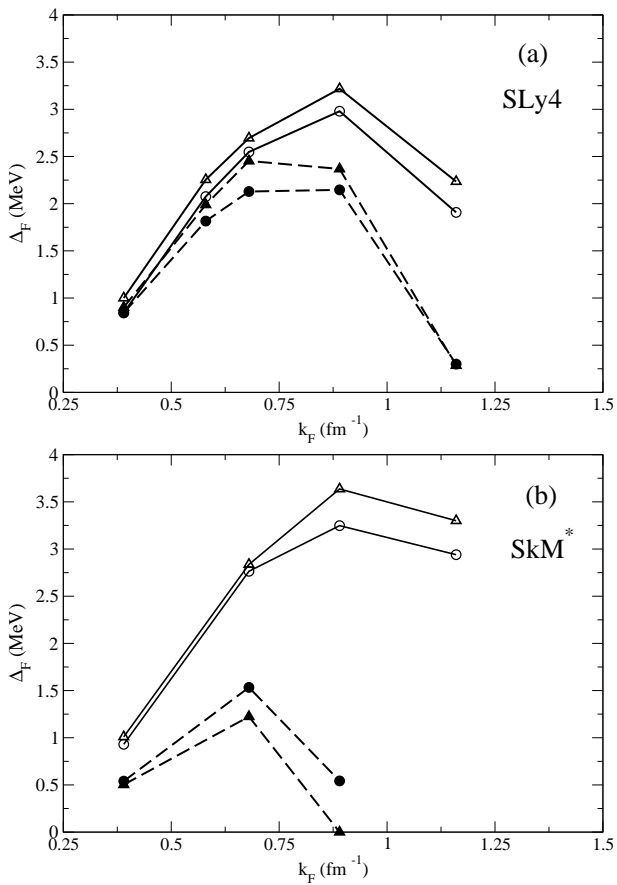

\centering
\includegraphics*[width=0.45\textwidth,angle=0]{pairing_gap}
\includegraphics*[width=0.45\textwidth,angle=0]{pairing_gap_SkmS}
\caption{Pairing gap calculated at the Fermi energy for the different WS cells.
Empty and filled triangles refer to the calculation without protons, performed with 
the Gogny interaction only (empty triangles)  and with the Gogny plus the induced interaction
(filled triangles), already shown in Fig.  \ref{IND eye: fig 60}. Empty and filled  
circles correspond to calculation with protons (nucleus placed at the center of the cell),
performed with  the Gogny interaction only (empty circles)  and with the Gogny plus the induced interaction
(filled circles). Mean field  and RPA modes are computed making use of the SLy4  (a) and of the SkM$^\ast$ (b) interactions.}
\label{IND eye: fig 50}
\end{figure}

We now turn our attention to the calculation including protons.
The pairing field obtained for all the computed cells is shown in Fig. \ref{IND eye: fig 50}. 
The circles refer to the calculation with the nucleus at the center of the cell,
while the triangles refer to the 
same gaps without nucleus already shown  in Fig. \ref{IND eye: fig 60}.

\begin{figure}[!ht]
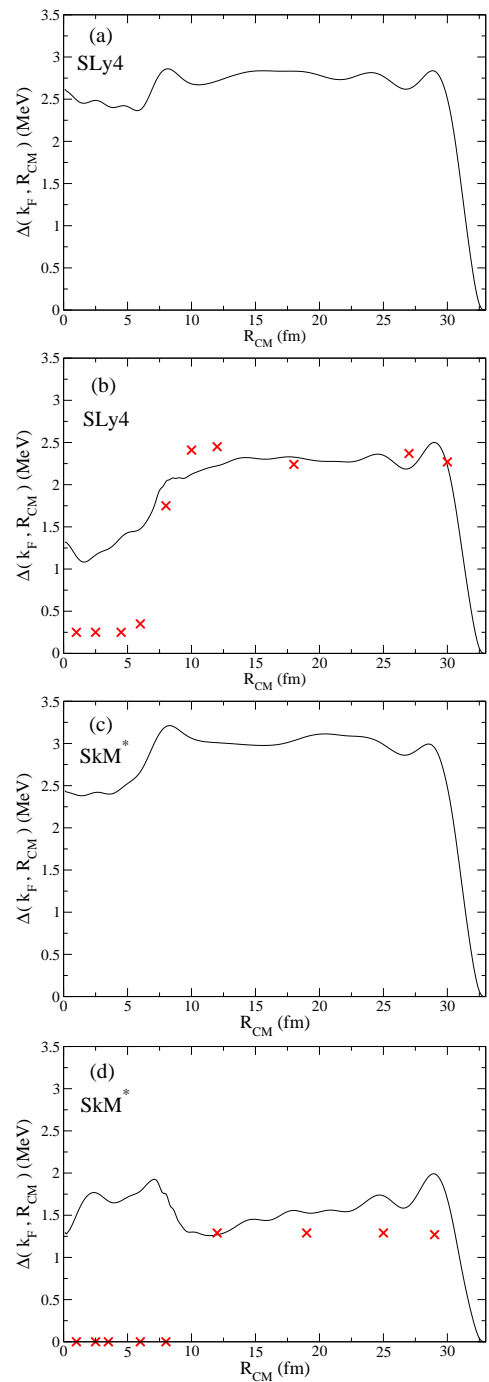

\centering
  \includegraphics*[width=0.35\textwidth,angle=0]{grafico_pairing_a_kf_1350Sn_G} 
  \includegraphics*[width=0.35\textwidth,angle=0]{grafico_pairing_a_kf_1350Sn_G_I_conpunti}\\
  \includegraphics*[width=0.35\textwidth,angle=0]{grafico_pairing_a_kf_1350Sn_G_SkmS} 
  \includegraphics*[width=0.35\textwidth,angle=0]{grafico_pairing_a_kf_1350Sn_G_I_SkmS_conpunti}   
\caption{
Pairing field $\Delta(k_F,R_{CM})$ for the cell 
\elem{1364}{Sn}, calculated with the SLy4 mean field and the Gogny pairing interaction (a),
with the SLy4 mean field and the Gogny+induced  pairing interaction (b),
with the SkM* mean field and the Gogny pairing interaction (c) and 
with the SkM* mean field and the Gogny+induced  pairing interaction (d).
The crosses in (b) and (d) are obtained dividing the pairing field shown in (a) and (c) respectively, 
by the quenching factor calculated in  uniform neutron matter.
%
%
}
\label{IND gap:fig 20}
\end{figure}

In Fig. \ref{IND eye: fig 50}(a)   we show the results obtained with the SLy4 interaction.
One can see that the presence of the nucleus reduces the pairing gap
by about 100-200 keV, both with and without the induced interaction.
The result without the induced interaction is similar to that obtained 
in ref. \cite{Pizzochero}, where it was shown that the pairing gap
is reduced by about 200 keV close to the Fermi energy. This effect was
attributed to the fact that pairing is quenched inside the nuclear volume, where the local
Fermi momentum is larger than in the outer neutron gas.
This can be seen in a clear way in Fig. \ref{IND gap:fig 20}(a) where we show
the spatial dependence of the pairing gap for the case with nucleus at $\rho_{n}=$0.01 fm$^{-3}$. 
This is obtained from our quantum calculation
as explained in  ref. \cite{Esbensen}: first we construct the non local pairing 
field $\Delta(\vec r_1, \vec r_2)$; then we perform a Fourier transform repect to the 
relative coordinate $\vec r_1- \vec r_2$, obtaining the gap as a function of the relative momentum $k$
and of the center of mass coordinate $R_{CM}$,
$\Delta(k,R_{CM})$ (we neglect the weak dependence on the angle between $\vec k$ and $\vec R_{CM}$); 
finally we define a local
pairing gap $\Delta_{loc}(R_{CM})=\Delta(k_F(R_{CM}),R_{CM})$, where $k_F(R_{CM})$ is the local 
Fermi momentum.
From this figure one observes that in fact the local pairing gap is reduced from 2.7 MeV far away
from the nucleus, to 2.4 MeV in the interior, and thus the mean value of the gap in the whole cell
is also reduced (the reduction would be stronger using the bare $v_{14}$ Argonne interaction instead of  the 
Gogny interaction, cf. Figs. 5 and 10 of ref. \cite{Pizzochero}).

Turning now to the calculations including the induced interaction, considering  
the results obtained in  neutron matter,
where the pairing gap goes to zero for $k_F>$ 1.2 fm$^{-1}$, one might expect that the 
pairing would almost vanish inside the nucleus (cf. crosses in  
Fig. \ref{IND gap:fig 20}(b)). Instead the actual local pairing gap resulting from the 
microscopic calculation (solid curve in Fig. \ref{IND gap:fig 20}(b))  shows that the gap inside the nucleus has a value of about 1.3 MeV,
which represents  a much smaller reduction. 
This is to be interpreted as a result 
of the enhancement of the attractive (negative) induced matrix elements due to the exchange
of nuclear collective surface modes, as it was discussed in the previous section 
(cf. also Fig.  \ref{NS ind:fig 30}).
The gap in the outer neutron gas remains  larger than in the interior of the nucleus, so that
the resulting average gap in the cell is somewhat smaller than in the case without the nucleus.

Considering now the results obtained with the SkM* interaction, one can observe two main differences
comparing Figs. \ref{IND eye: fig 50} (a) and (b):
first, as already discussed, the screening effect due to the interaction is much stronger 
than in the SLy4 case; secondly,  the effect of the nucleus now leads to an increase
of $\Delta_F$ when the induced interaction is included. 
This is due to the fact that, as we have seen above, the SkM* interaction leads to  a much larger screening 
compared to SLy4 in uniform matter at low density. In fact, the induced interaction 
leads to rather similar values of the gap inside the nucleus for the two forces,
while spin modes reduce  the gap in the external  neutron gas much more with  the SkM* than with the 
SLy4 force (cf. Fig.  \ref{IND gap:fig 20}(b) and (d)). 
In summary, the action of the induced interaction acts in a similar way for SkM* and SLy4 in the interior of
the nucleus,  where the action of spin modes (which at saturation density is about the same
for the two forces, cf. Fig. \ref{gogny vs residual}) is strongly counteracted by surface modes; while in the outer,
low-density neutron gas the induced interaction associated with  SkM* suppresses the gap much more
effectively than in the SLy4 case.

\section{Conclusions}
The effect of the induced interaction arising from the exchange of 
density and spin modes of the system results in a quenching
of the neutron pairing gap, both in the case of uniform neutron matter 
and in the case of the inner crust of a neutron star.


The pairing gap obtained is found to be strongly dependent on the effective nucleon-nucleon interaction employed to determine
the single-particle properties of the system and its linear response (we performed the calculation making use of SLy4 and SkM$^\ast$ Skyrme interactions).
A much stronger suppression of the gap is obtained in uniform neutron matter when the 
 SkM$^\ast$ interaction is used, due to the stronger residual interaction associated with this NN effective force
and to the different balance between spin and density modes.

When a nuclear cluster is present in the cell, the local Fermi momentum increases and the pairing gap 
calculated with the bare interaction is reduced in the interior of the nucleus. 
The induced interaction quenches the gap both inside in the nucleus and in the outer neutron sea:
the relative amount depends on the interaction, but the reduction inside the nucleus is much smaller,
than that occurring in neutron matter at the corresponding Fermi momentum.

\section{Acknowledgements}
We thank G. Gori for fruitful discussions and for the great help
he gave us. G. Col\`o provided the code employed in  the RPA calculation.
The support of the
supercomputing group at the Consorzio Interuniversitario Lombardo Elaborazione Automatica (CILEA),
Segrate, Italy, has been essential for the calculations.
F.B.  aknowledges partial support from the Spanish Education and Science Ministery projects FPA2006-13807-c02-01,
FIS2005-01105 and INFN08-33. The work of S.B. is supported in part by
the Natural Sciences and Engineering Research Council of Canada (NSERC). 
TRIUMF receives federal funding via a contribution agreement through the National Research Council of Canada.

\vspace{1cm}

\section{Appendix A}\label{App A}

The detailed calculations of the pairing gap performed in the WS cell without protons can be checked  against a much simpler estimate in neutron matter, using a similar 
assumption for  the residual interaction, which  is taken into account within the 
Landau approximation, neglecting the momentum dependence of the Landau parameters.
However, the response function will also be estimated within the Landau approximation,
while in the microscopic calculation in the WS cell the full RPA response has been considered.    
The matrix elements of the induced interaction in infinite neutron matter
are obtained summing the contributions $v_{ind}^{S=0}$ and $v_{ind}^{S=1}$
in the $S=0$ and $S=1$ channels. They
are obtained multiplying the square of the Landau parameters $C_F$ (density modes)
and $C_G$ (spin modes) by the  integral of the RPA response function $R^{S=0}(q)$
(density modes) and $R^{S=1}(q)$ (spin modes) \cite{Viverit,Schulze}.
For density modes one has
\begin{equation}
v_{ind}^{S=0} (k_1,k_2) = \frac{C_F^2(k_F)}{k_1k_2} \int_{q_{min}}^{q_{max}} dq q R_{S=0}(q),
\end{equation}
where $k_1$ and $k_2$ are the momenta of the states exchanging the vibrations. 
The integration limits
are equal to $q_{min}=|k_1 -k_2|$ and to $q_{max}= k_1 + k_2$.
One has 
\begin{equation}
R^{S=0}(q) = -\frac{1}{2} \frac{N(0) L(q)}{1 - C_F(k_F) L(q)}
\label{RPAnm}
\end{equation}
where $L(q)$ is the Lindhardt function and  $N(0)$ is the density of single-particle
states at the Fermi energy, $N(0)= \frac{k_f m^*}{\hbar^2\pi^2}$.
The contribution from spin modes is correspondingly obtained from 
$R^{S=1}$, where 
\begin{equation}
R^{S=1}(q) = -\frac{1}{2} \frac{N(0) L(q)}{1 - C_G(k_F) L(q)},
\end{equation}
and from
\begin{equation}
v_{ind}^{S=1}(k_1,k_2) = - \frac{3 C_G^2(k_F)}{k_1 k_2} \int_{q_{min}}^{q_{max}} dq q R_{S=1}(q),
\end{equation}
where the factor 3 is associated with the three spin projections in the $S=1$
channel. 

The pairing gap is then calculated solving the BCS equations with the complete
interaction, $v_{tot}(k_1,k_2) = v_{bare} + v_{ind}^{S=0}+v_{ind}^{S=1}$.  
In Fig. \ref{IND eye: fig 60}
we have compared  these simple expressions with the microscopic calculations
in the Wigner cell without nuclei.





\bibliographystyle{apsrev}

\begin{thebibliography}{99}


 \bibitem{Pet.Rav.:1995} C.J.Pethick and D.G.Ravenhall,
 Ann. Rev. Nucl. Part. Sci. {\bf 45} (1995) 429
\bibitem{Lombardo} U. Lombardo and H.-J. Schulze, in {\it Physics of neutron star interiors},
eds. D. Blaschke, N.K. Glendenning and A. Sedrakian, Springer (2001), p.31
\bibitem{Cao} L.G. Cao, U. Lombardo and P. Schuck, Phys. Rev. {\bf C74}(2006)064301.
\bibitem{Gori2} G.Gori, F. Ramponi, F. Barranco, R.A. Broglia, G. Col\`o and E. Vigezzi,
Phys. Rev. {\bf C 72} (2005)011302(R).
\bibitem{Gori1} G.Gori, F. Ramponi, F. Barranco, R.A. Broglia, G. Col\`o, D. Sarchi and E. Vigezzi,
Nucl. Phys. {\bf A731} (2004)401
\bibitem{NegVau} J. Negele and D. Vautherin, Nucl. Phys. {\bf A207} (1973) 
\bibitem{Krivine} H. Krivine, J. Treiner and O. Bohigas, Nucl. Phys. {\bf A336} (1984) 155.
\bibitem{Cha.ea.:1998} E. Chabanat, P.Bonche, P.Haensel, J.Meyer and R.Schaeffer,
 Nucl. Phys. {\bf A635} (1998) 231
\bibitem{Montani} F. Montani, C. May and H. M\"uther, Phys. Rev. {\bf C69} (2004) 065801.
\bibitem{Sandulescu}  N. Sandulescu, N. Van Giai
and R.J. Liotta, Phys. Rev. {\bf C69} (2004) 045802;
N Sandulescu, Phys. Rev. {\bf C70} (2004) 025801.
\bibitem{Pizzochero} P.M. Pizzochero, F. Barranco, R.A. Broglia and E. Vigezzi, 
Astrophys. Jou. {\bf 569} (2002) 381.
\bibitem{Vangiai} C. Garcia-Recio, J. Navarro, N. Van Giai and L. Salcedo, 
Ann. Phys, {\bf 214} (1992) 293.
 \bibitem{Blo.Hor:1955} C.Bloch and J.Horowitz,
 Nucl. Phys. {\bf 8} (1955) 1991
\bibitem{Dus.Lio:1971} G.Dussel and R.Liotta,
 Phys. Lett. {\bf 37} (1971) 477
\bibitem{Bes.ea:1976} D. R. Bes, R.A.Broglia, G. G. Dussel, R. J. Liotta and H. M. Sofia,
 Nucl. Phys. {\bf A260} (1976) 1
\bibitem{Bes.ea:1977} D.R.Bes, G.Dussel and H.Sofia,
 Am. J. of Phys.{\bf 55 N.2} (1977) 81
\bibitem{Bar.ea:1999} F.Barranco, R.A.Broglia, G.Gori, E.Vigezzi, P.F.Bortignon and J.Terasaki,
 Phys. Rev. Lett. {\bf 83} (1999) 2147
\bibitem{Ter.ea:2002b} J.Terasaki, F.Barranco, R.A.Broglia, E.Vigezzi, P.F.Bortignon,
 Progr. Theor. Phys. {\bf 108} (2002) 495
\bibitem{Bar.ea:2004b} F.Barranco, R.A.Broglia, G.Col\`o, G.Gori, E.Vigezzi, P.F.Bortignon,
 Eur. J. of Phys. {\bf 21} (2004) 57
\bibitem{Dec.Gog:1980} J.Decharg\'e and D.Gogny,
 Phys. Rev. C {\bf 21} (1980) 1568
\bibitem{Schuck} F.Barranco, P.F. Bortignon, R.A.Broglia, G. Col\`o, P. Schuck, E.Vigezzi and
X. Vi\~nas, Phys. Rev. {\bf C72} (2005) 054314.
\bibitem{Esbensen} F.Barranco, R.A.Broglia, H. Esbensen and  E.Vigezzi,
Phys. Rev {\bf C58} (1998) 1257.
\bibitem{Viverit} H. Heiselberg, C.J. Pethick, H. Smith and L. Viverit, Phys. Rev. Lett.
{\bf 85} (2000) 2418.
\bibitem{Schulze} H.-J. Schulze, J. Cugnon, A. Lejeune, M. Baldo and 
U. Lombardo, Phys. Lett. {\bf B375} (1996) 1.
\end{thebibliography}

\end{document}